\documentclass[11pt,twoside,a4paper,cmspaper,final,collab]{cms-tdr}
\def\svnVersion{89d9939}\def\svnDate{2026/07/01}\def\cmsCernNoTag{CERN-EP-2026-032}\def\cmsCernDate{\today}\def\cmsMessage{Published in Physics Letters B as \href{http://dx.doi.org/10.1016/j.physletb.2026.140679}{\doi{10.1016/j.physletb.2026.140679}.}}
\begin{document}\cmsNoteHeader{HIN-25-014}

\newcommand{\pp}{\ensuremath{\Pp\Pp}\xspace}
\newcommand{\pPb}{\ensuremath{\Pp\text{Pb}}\xspace}
\newcommand{\PbPb}{\ensuremath{\text{PbPb}}\xspace}
\newcommand{\OO}{\ensuremath{\text{OO}}\xspace}
\newcommand{\NeNe}{\ensuremath{\text{NeNe}}\xspace}
\newcommand{\XeXe}{\ensuremath{\text{XeXe}}\xspace}
\newcommand{\cubicA}{\ensuremath{A^{1/3}}\xspace}
\newcommand{\A}{\ensuremath{A}\xspace}
\newcommand{\RAA}{\ensuremath{R_{\text{AA}}}\xspace}
\newcommand{\qHat}{\ensuremath{\hat{q}}\xspace}
\newcommand{\dSigdPtAA}{\ensuremath{\frac{\rd \sigma^{\text{AA}}}{\rd \pt}}\xspace}
\newcommand{\dSigdPtpp}{\ensuremath{\frac{\rd \sigma^{\pp}}{\rd \pt}}\xspace}
\newcommand{\dNdEta}{\ensuremath{\rd N_{\text{ch}}/\rd\eta}\xspace}
\newcommand{\bini}{\ensuremath{i}\xspace}
\newcommand{\ptwidthi}{\ensuremath{\Delta p_{\mathrm{T},\bini}}\xspace}
\newcommand{\ptcenteri}{\ensuremath{p_{\mathrm{T},\bini}}\xspace}
\newcommand{\ptwidthMerged}{\ensuremath{\Delta p_{\mathrm{T}}^{\,\text{merged}}}\xspace}
\newcommand{\ptcenterMerged}{\ensuremath{p_{\mathrm{T}}^{\,\text{merged}}}\xspace}
\newcommand{\sigPbPb}{\ensuremath{\sigma^{\mathrm{PbPb}}}\xspace}
\newcommand{\sigBin}{\ensuremath{\sigma^{\mathrm{PbPb}}_{\text{bin},\bini}}\xspace}
\newcommand{\sigBinStat}{\ensuremath{\sigma^{\mathrm{PbPb}}_{\text{bin, stat},\bini}}\xspace}
\newcommand{\sigBinSyst}{\ensuremath{\sigma^{\mathrm{PbPb}}_{\text{bin, syst},\bini}}\xspace}
\newcommand{\sigMerged}{\ensuremath{\sigma^{\mathrm{PbPb}}_{\text{merged}}}\xspace}
\newcommand{\sigMergedStat}{\ensuremath{\sigma^{\mathrm{PbPb}}_{\text{stat, merged}}}\xspace}
\newcommand{\sigMergedSyst}{\ensuremath{\sigma^{\mathrm{PbPb}}_{\text{syst, merged}}}\xspace}

\newlength{\cmsFigWidth}

\ifthenelse{\boolean{cms@external}}
{   
\renewcommand{\appendixname}{Appendix}
\newcommand{\addmatter}[2]{supplementary material #2}
\setlength{\cmsFigWidth}{\columnwidth}
}
{
\newcommand{\addmatter}[2]{\appendixname~\ref{#1}}
\setlength{\cmsFigWidth}{0.9\linewidth}
}

\cmsNoteHeader{HIN-25-014}

\title{System-size dependence of charged-particle suppression in ultrarelativistic nucleus-nucleus collisions}

\date{\today}

\abstract{
High-energy partons lose energy while propagating through the hot, strongly interacting medium produced in ultrarelativistic nucleus-nucleus collisions, leading to a suppression of particle production at high transverse momentum (\pt). The dependence of this energy loss on the size of the colliding nuclear system has yet to be firmly established experimentally. This Letter presents a systematic study of charged-particle suppression across four different nucleus-nucleus collision systems using nuclear modification factors (\RAA) measured by the CMS Collaboration at the CERN LHC. Previous CMS measurements of \RAA in oxygen-oxygen, xenon-xenon, and lead-lead collisions are recast with identical \pt intervals and are complemented by the first measurement of the charged-particle \RAA in neon-neon collisions at $\sqrtsNN = 5.36\TeV$. The neon-neon data correspond to an integrated luminosity of 0.76\nbinv. The \RAA in all collision systems examined show similar qualitative trends as a function of \pt, but have a magnitude which is ordered with the nucleon number \A. The \RAA feature a downward slope at low \pt, a local minimum at around 5--7\GeV, and an upward slope with increasing \pt. The \RAA are also compared in terms of \cubicA, which is proportional to the nuclear radius. Models including only initial-state nuclear effects fail to reproduce the observed trends, whereas energy loss models reproduce the trends in the region $\pt > 9.6\GeV$.
}

\hypersetup{
pdfauthor={CMS Collaboration},
pdftitle={System-size dependence of charged-particle suppression in ultrarelativistic nucleus-nucleus collisions},
pdfsubject={CMS},
pdfkeywords={CMS, light ions, heavy ions, jet quenching, quark-gluon plasma}}

\maketitle 

\section{Introduction}

In ultrarelativistic heavy ion collisions, a hot and deconfined state of strongly interacting matter, known as quark-gluon plasma (QGP), is formed~\cite{Busza:2018rrf,CMS:2024krd}. High-energy partons produced in initial hard scatterings traverse the QGP and probe its short-distance structure~\cite{dEnterria:2009xfs,Cunqueiro:2021wls,Apolinario:2022vzg}. As a consequence of their interactions with the surrounding medium, partons lose energy while propagating through it~\cite{Gyulassy:1990ye,Bjorken:1982tu}. A characteristic experimental signature of parton energy loss in heavy ion collisions is the suppression of high transverse momentum (\pt) particle yields relative to those measured in a reference proton-proton (pp) collision system.

The magnitude of parton energy loss depends on the presence or absence of QGP formation and, if such a medium is formed, on its properties, including the temperature and transport properties characterizing parton-medium interactions. In the presence of the QGP, the amount of parton energy loss is expected to depend primarily on the in-medium path length traversed by the parton, which is strongly influenced by the size of the colliding nuclear system. In this context, the system size refers to the spatial extent of the nuclear overlap region, which is largely determined by the radii of the colliding nuclei. Energy loss proceeds via a combination of different mechanisms, \eg, collisional or radiative processes, which have different path-length dependences.  Variations in the local temperature, the density, and the dynamic expansion of the medium also affect the path-length dependence of energy loss.  Thus, the exact scaling of parton energy loss with system size is a theoretically challenging topic, and experimental input is needed to provide additional constraints. To date, studies of the system-size dependence have largely relied on centrality-selected observables in heavy ion collisions~\cite{CMS:2025jbv,CMS:2018yyx,ATLAS:2022iyq,ALICE:2017svf,CMS:2016xef,ALICE:2014nqx,ATLAS:2022kqu}. Centrality is used to quantify the geometric overlap of the two colliding nuclei and is based on the percentile of the total inelastic nucleus-nucleus (AA) cross section, with lower (higher) percentiles corresponding to more central (more peripheral) collisions. These measurements predominantly explore variations in collision geometry within a single nuclear system, rather than comparisons across different systems. They are affected by selection and geometry biases inherent to the centrality determination, particularly in peripheral events~\cite{Loizides_2017, CMS:2018yyx, CMS:2021kvd}. As a result, a quantitative and unbiased characterization of the system-size dependence of parton energy loss remains an open challenge, limiting the development of a unified theoretical description of nuclear suppression phenomena.

A characterization of nuclear suppression effects based instead on the nucleon number \A provides a model-independent measure of the overall system size and a less biased approach to study the system-size dependence of parton energy loss. Recent light ion collision data, i.e., from oxygen-oxygen (OO) and neon-neon (NeNe) collisions~\cite{Brewer:2021kiv,Bruce:2021hjk,AlemanyFernandez:2025ixd}, have provided a new avenue to probe the onset of parton energy loss effects in small nuclear collision systems. These data circumvent the aforementioned issues of event selection and geometric biases and complement the existing measurements in xenon-xenon (XeXe) and lead-lead (PbPb) collisions. In addition, high-\pt particle production preferentially selects more head-on central collisions, further reducing the range of collision geometries sampled by the measurement within a given collision system.

Nuclear suppression effects from parton energy loss can be quantified using the charged-particle nuclear modification factor, \RAA. For centrality-integrated (minimum bias) AA collisions, this factor can be defined as
\begin{equation}
\RAA = \frac{1}{A^2}
\left[
\left(\dSigdPtAA\right)\bigg{/}
\left(\dSigdPtpp\right)
\right],
\label{eq1}
\end{equation}
where $\sigma^{\text{AA}}$ ($\sigma^{\pp}$) denotes the charged-particle production cross section in nucleus-nucleus (\pp) collisions. The $A^2$ normalization originates from having \A nucleons in each nucleus and therefore constitutes an exact quantity, avoiding model-dependent estimates of the nuclear overlap~\cite{dEnterria:2020dwq,Loizides:2016djv}. Deviations of \RAA from unity indicate modifications of particle production arising from nuclear effects, including both initial- and final-state interactions. Owing to the steeply falling \pt spectrum, even modest amounts of energy loss can lead to a sizable suppression of the observed particle yields, making \RAA a particularly sensitive probe of medium-induced energy loss.

Recently, the CMS Collaboration reported the discovery of suppressed charged-particle production in inclusive \OO collisions at a center-of-mass energy per nucleon pair $\sqrtsNN = 5.36\TeV$ by measuring the charged-particle \RAA~\cite{CMS:2025bta}. When combined with existing centrality-integrated \PbPb measurements at $\sqrtsNN = 5.02\TeV$~\cite{CMS:2016xef} and \XeXe measurements at 5.44\TeV~\cite{CMS:2018yyx}, these data provide a unique opportunity for an experimental constraint of parton energy loss effects versus the nuclear system size. The \XeXe measurement uses the 80\% most central collisions and is therefore expected to provide a close approximation to the inclusive \RAA.

Motivated by these developments, this Letter investigates the system-size dependence of high-\pt charged-particle production in AA collisions at the CERN LHC. In addition to existing charged-particle \RAA measurements, this analysis presents the first measurement of the charged-particle \RAA in \NeNe collisions at $\sqrtsNN = 5.36\TeV$, adding an additional nuclear system having a size between that of \OO and \XeXe collisions. The four collision systems are presented in identical charged-particle \pt intervals to facilitate the systematic evaluation of the charged-particle \RAA as functions of \pt and \A. Tabulated results are provided in the HEPData record for this analysis~\cite{hepdata}.

\section{Detector description}\label{sec:CMSdetectorDescription}

The CMS apparatus~\cite{CMS:2008xjf,CMS:2023gfb} is a multipurpose, nearly hermetic detector, designed to trigger on~\cite{CMS:2020cmk,CMS:2016ngn,CMS:2024aqx} and identify electrons, muons, photons, and hadrons~\cite{CMS:2020uim,CMS:2018rym,CMS:2014pgm}. A global ``particle-flow'' algorithm~\cite{CMS:2017yfk} aims to reconstruct all individual particles in an event, combining information provided by the all-silicon tracker and by the crystal electromagnetic and brass-scintillator hadron calorimeters, operating inside a 3.8\unit{T} superconducting solenoid, with data from the gas-ionization muon detectors embedded in the flux-return yoke outside the solenoid. The silicon tracker~\cite{Phase1Pixel} measures charged particles up to pseudorapidity $\abs{\eta} = 3.0$ with typical resolutions of 2.0\% in \pt and 10--75\mum in the transverse impact parameter~\cite{DP-2020-049} for nonisolated particles of $1 < \pt < 100\GeV$. The measurement is restricted to $\abs{\eta} < 1$, in order to be able to compare with previous measurements. This is also a $\eta$ interval where the \pt resolution and track reconstruction have optimal performance. The hadron forward (HF) calorimeter uses steel as an absorber and quartz fibers as the sensitive material. The two halves of the HF are located 11.2\unit{m} from the interaction region, one on each end from the interaction region, and together they provide coverage in the range $3.0 < \abs{\eta} < 5.2$.

\section{Charged-particle \texorpdfstring{\RAA}{RAA} in NeNe collisions}\label{sec:NeNeRAA}

In general, the \NeNe measurement strategy follows the one devised for the charged-particle \RAA measurement in \OO collisions in Ref.~\cite{CMS:2025bta}. The same \pp reference from Ref.~\cite{CMS:2025bta} is used to evaluate the charged-particle \RAA in \NeNe collisions.

The \NeNe data were collected in 2025 using the same online selection as that employed in the \OO analysis, based on energy deposits in the HF calorimeters in coincidence with colliding bunches. The average number of interactions per bunch crossing, typically referred to as pileup, was approximately 0.13. The total integrated luminosity used in this analysis is $0.76 \pm 0.04\nbinv$. This value is obtained by combining two independent luminosity measurements. The first method is based on a preliminary analysis of van der Meer scan data~\cite{vanderMeer:296752,CMS:2025rzq}. An independent luminosity determination was performed by comparing the observed \PZ boson yields to a well-calibrated reference data set~\cite{CMS:2023pad}. The luminosities obtained with the two approaches are in good agreement; their central values differ by only 4.1\%.

The \NeNe \RAA analysis is reported for primary charged particles, defined as charged particles with a proper decay length $c\tau>1\unit{cm}$. Particles that result from secondary decays are not considered primary if the parent particle is classified as primary. Charged particles are reconstructed in the silicon tracker. To reduce the contributions from secondary tracks, tracks are required to have a distance of closest approach (DCA) to any primary vertex smaller than three times the corresponding DCA uncertainty in both the transverse and longitudinal directions. Track reconstruction, selection criteria, and kinematic requirements ($\abs{\eta}<1$ and $3.0<\pt<103.6\GeV$) are the same as those used in the \OO measurement and in the corresponding \pp reference analysis~\cite{CMS:2025bta}. These selections ensure minimal bias from online and offline event requirements and maximize the cancellation of reconstruction-related uncertainties in the \RAA ratio. Restricting the measurement to $\pt < 103.6\GeV$ improves control over contributions from misreconstructed tracks, which become increasingly challenging to constrain in the dense track environment of high-\pt jet showers.

Event selection requirements are chosen to efficiently select minimum bias \NeNe collisions, while suppressing backgrounds from ultraperipheral~\cite{Baltz:2007kq} and beam-related interactions. Offline event selections follow exactly those of the \OO analysis~\cite{CMS:2025bta}, and rely on coincidence signals that produce energy deposits above 13\GeV in one or more detector elements on each half of the HF. The presence of at least one reconstructed primary vertex~\cite{CMS:2014pgm} is also required. The resulting inefficiency arising from the event selection requirements for minimum bias events is below 1\% for the fiducial region of the charged particles considered in this analysis. This inefficiency, particularly its impact on the charged particle \pt spectrum, is taken into account with the corrections to the primary charged-particle level \pt spectrum. Ion beam contamination, such as that from transmutation from the electromagnetic dissociation of $^{20}$Ne~\cite{Pshenichnov:1997un}, was found to be negligible based on the time dependence of distributions characterizing the event activity~\cite{Das:2025rqb}.

Simulated Monte Carlo events are employed to determine tracking efficiency and acceptance corrections, to assess the impact of momentum resolution and residual background contamination, and to validate the stability of the correction procedures. Minimum bias \NeNe collisions are simulated using the \HIJING v1.383 generator~\cite{Wang:1991hta}. High-\pt charged-particle production is modeled by embedding \PYTHIA v8.311~\cite{Bierlich:2022pfr} events with the CP5 tune~\cite{CMS:2019csb} into \HIJING events, forming composite events that combine a hard scattering process with particle production from the underlying event~\cite{CMS:2025bta}. Potential contamination from ultraperipheral collisions~\cite{Baltz:2007kq} is studied using the \textsc{STARlight} v3.3.0 generator~\cite{Klein:2016yzr} and is found to be negligible for the phase-space region reported in this measurement. The detector response is simulated using \GEANTfour~\cite{Agostinelli:2002hh}. After event generation, simulated events are weighted on an event-by-event basis such that the distribution of the position along the beam axis of reconstructed primary interaction vertices matches the one observed in data. Differences between data and simulation in the charged-particle species composition, such as the fractional composition of charged pions, kaons, protons, and strange hadrons, are corrected using the same procedure as in the \OO analysis~\cite{CMS:2025bta}. The procedure uses as input the \pt-dependent species fractions measured by the ALICE experiment in \pp collisions~\cite{ALICE:2020jsh,ALICE:2016dei}, which are further rescaled guided by the measured hadrochemistry evolution with the average charged-particle density (\dNdEta)~\cite{ALICE:2019hno,ALICE:2016fzo} in order to match the expected composition for the \NeNe collision system. The measurements in Refs.~\cite{ALICE:2019hno,ALICE:2016fzo} indicate that the particle species composition is governed by particle activity, independently of the collision system. For this correction, we assume a reference value of $\dNdEta=50$ for \NeNe collisions, determined from the measured $\dNdEta=40$ in minimum bias \OO collisions~\cite{CMS-PAS-HIN-25-010} scaled to the \NeNe system using the Glauber calculations in Ref.~\cite{Loizides:2025ule} as input. The \pt-dependent correction factors are similar, to the per-mille level, to the ones from \OO, as expected from the observed evolution of the particle species for those values of \dNdEta~\cite{ALICE:2019hno,ALICE:2016fzo}.

\begin{figure}[htb!]
    \centering
    \includegraphics[width=\cmsFigWidth]{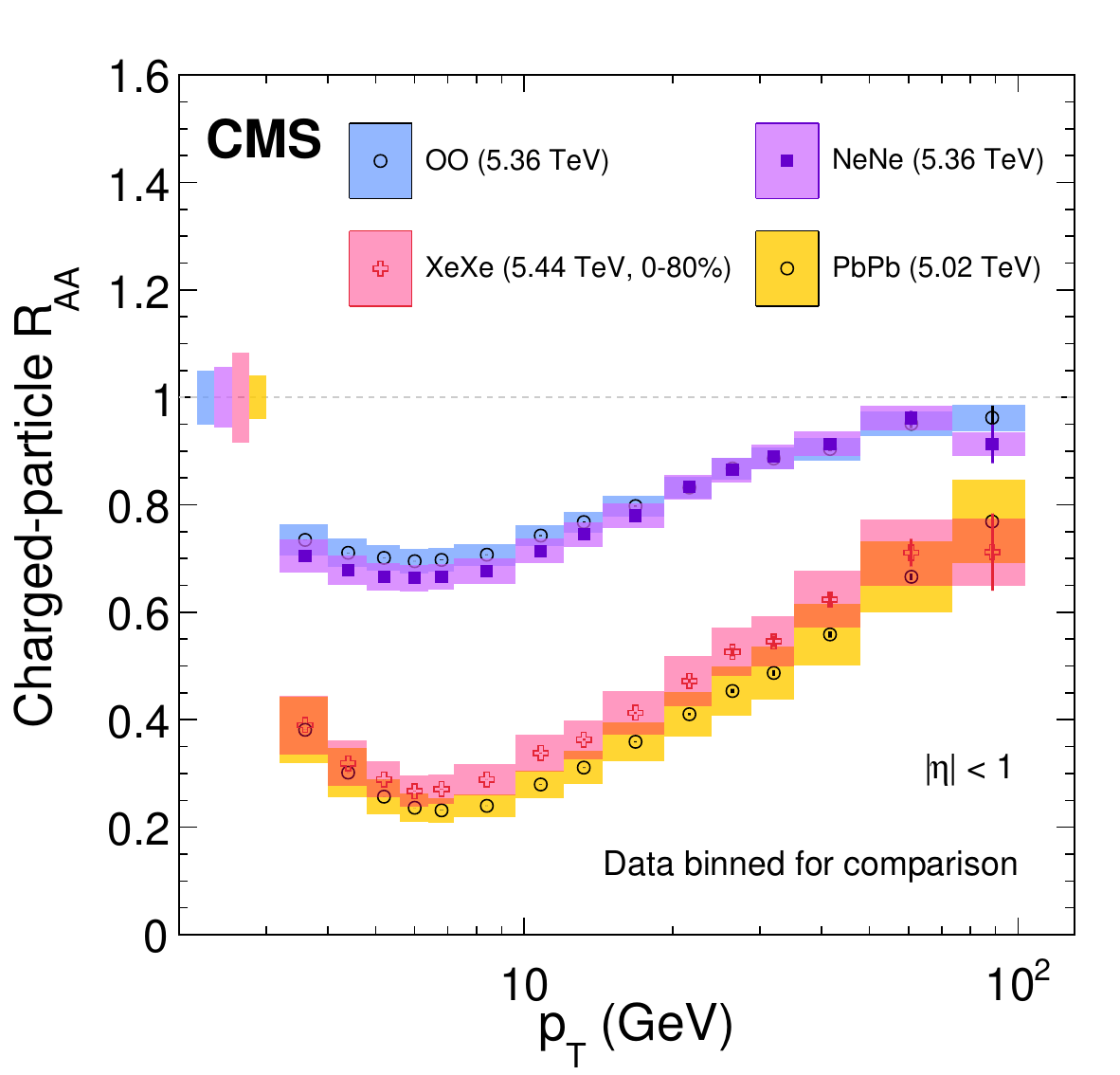}
	\caption{Charged-particle \RAA values versus \pt measured in \NeNe collisions at 5.36\TeV compared to previous measurements in centrality-integrated \OO collisions at 5.36\TeV~\cite{CMS:2025bta}, 0--80\% centrality \XeXe at 5.44\TeV~\cite{CMS:2018yyx}, and minimum bias \PbPb collisions at 5.02\TeV~\cite{CMS:2016xef}. The vertical error bars represent statistical uncertainties and the boxes represent systematic uncertainties. Global normalization uncertainties are not included directly in the data markers error bands, but they are instead represented by the various boxes on the left around unity. For this compilation plot, a uniform \pt binning scheme is adopted to ensure a consistent comparison across collision systems, as described in the text. The same results with finer binning can be found in \addmatter{app:nominal}{Section~3}.}
    \label{fig:fig1}
\end{figure}

Systematic uncertainties are evaluated through dedicated variations of the relevant correction inputs, following the same procedures as in the \OO \RAA measurement~\cite{CMS:2025bta}. The uncertainty on the particle species inputs from \pp collisions is obtained by propagating the experimental uncertainties~\cite{ALICE:2020jsh,ALICE:2016dei} on the measured species fractions versus \pt; this effect cancels in the \RAA ratio but affects the charged-particle \pt spectrum. The particle species extrapolation uncertainty is determined by varying the assumed particle composition resulting from variations in the \dNdEta-based extrapolation for central \NeNe collisions. This uncertainty affects the spectrum and is propagated directly to the \RAA. The charged-particle \pt resolution uncertainty is propagated by applying an additional Gaussian smearing of 1\% to the data; this uncertainty affects the charged-particle \pt spectrum but cancels in the \RAA. The luminosity uncertainty for \NeNe collisions, 4.8\%, is assumed to be uncorrelated from the luminosity uncertainty of 3.0\% for \pp collisions~\cite{CMS:2025bta}, amounting to a global normalization uncertainty of 5.7\% in the \RAA observable. The tracking efficiency uncertainty is taken as a 2.1\% \pt-independent uncertainty~\cite{CMS-PAS-TRK-10-002}, fully correlated across \pt bins. This contribution cancels in the \RAA ratio but is present for the charged-particle \pt spectrum. Data-simulation differences in the rejection of secondary tracks are evaluated by varying the rejection selections, deriving new tracking efficiency tables for each variation, and taking the spread of the corrected yields as the corresponding systematic uncertainty. The DCA significance requirement is varied between two and five times its uncertainty. The resulting uncertainty on the spectrum cancels partially in the \RAA, and is under 3.5\% because of partial cancellations. The systematic uncertainties are combined in quadrature. The corrected charged-particle \pt spectrum can be found in \addmatter{sec:spectrum}{Section~1}. The resulting \RAA for \NeNe collisions is presented in Fig.~\ref{fig:fig1} together with other systems, following the \pt binning scheme discussed next.

\section{Charged-particle \texorpdfstring{\RAA}{RAA} dependence on \texorpdfstring{\pt}{pT} and \texorpdfstring{\A}{A}}\label{sec:chargedParticleRAA_results}

To enable a consistent system-size comparison, the \RAA measurements in \OO, \NeNe, \XeXe, and \PbPb collisions ($\A=16$, 20, 129, and 208) are computed using a common \pt binning scheme corresponding to that of the published \XeXe measurement~\cite{CMS:2018yyx}, which has the coarsest binning. The study includes CMS results from minimum bias \OO collisions at 5.36\TeV~\cite{CMS:2025bta}, minimum bias \NeNe collisions at 5.36\TeV, \XeXe collisions at 5.44\TeV in the 0--80\% centrality range~\cite{CMS:2018yyx}, and minimum bias \PbPb collisions at 5.02\TeV~\cite{CMS:2016xef}. The minor differences in collision energy are not expected to strongly affect the overall conclusions that can be drawn in the comparison of different systems, since comparisons of charged-particle \RAA measurements at 2.76 and 5.02\TeV show no strong dependence on collision energy at the \TeV scale~\cite{CMS:2016xef}. To recompute the \PbPb \RAA from the published data, the \PbPb and 5.02\TeV \pp spectra are rebinned by merging adjacent \pt intervals using \pt-weighted averages to preserve the underlying spectral shapes, with statistical uncertainties combined in quadrature and systematic uncertainties combined linearly assuming full bin-to-bin correlation, while normalization uncertainties remain unchanged. The same procedure is applied consistently to the 5.02\TeV \pp reference. The \OO and \NeNe \RAA measurements are repeated with the same binning
scheme using the original data sets---no assumptions are needed regarding the combination of adjacent \pt intervals for these data. Further details of the binning procedure are provided in \addmatter{app:binning}{Section~2}, and the \RAA results with the high-granularity binning format can be found in \addmatter{app:nominal}{Section~3}.

\begin{figure}[ht!]
    \centering    \includegraphics[width=\cmsFigWidth]{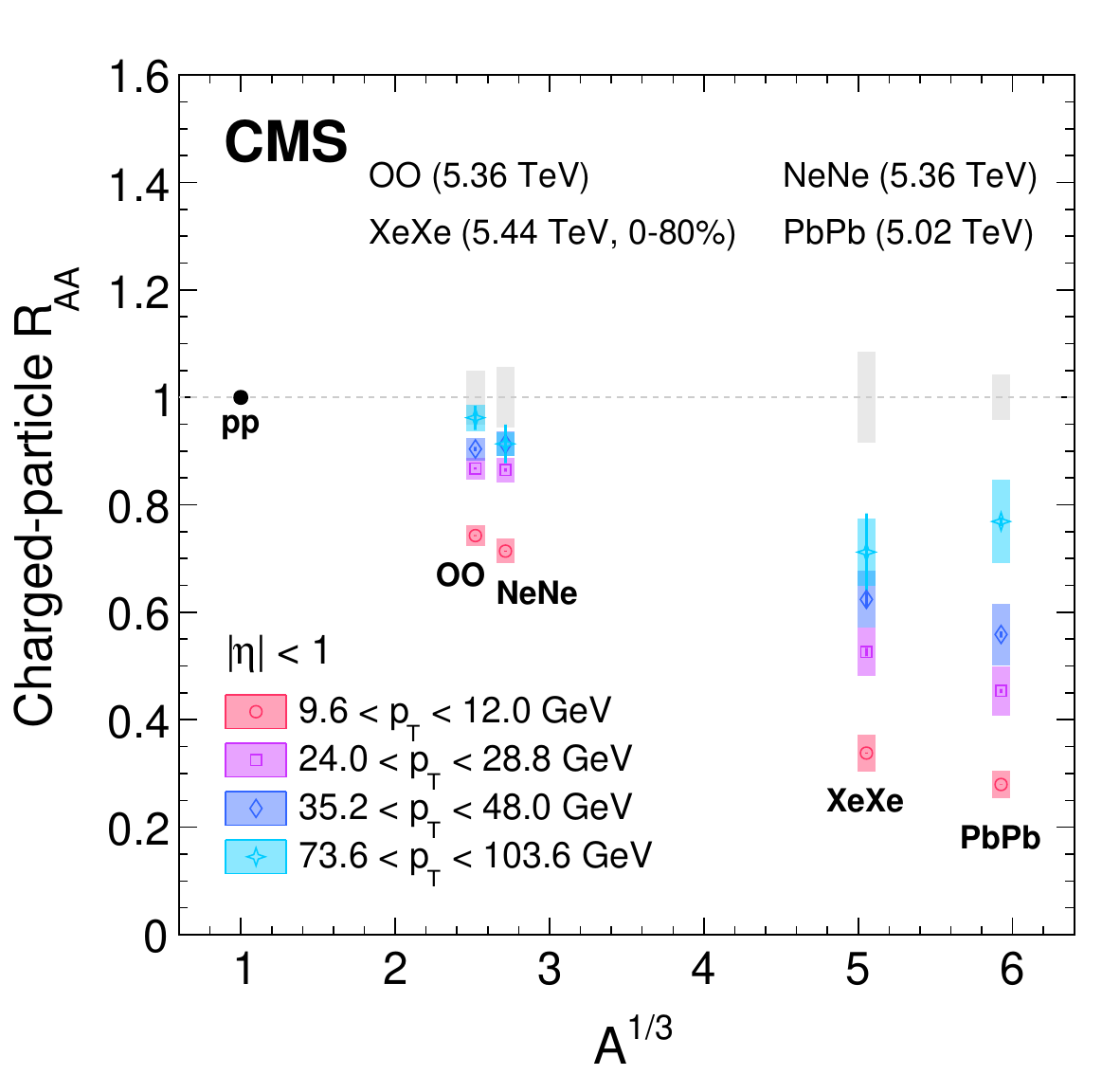}
    \caption{Charged-particle \RAA values, in intervals of \pt, versus \cubicA of the AA colliding system. The open markers represent the measured \RAA, with the vertical bars representing the statistical uncertainty, and the vertical band representing the experimental uncertainty. Global normalization uncertainties for each data set are shown in the light-gray boxes around unity. For visualization purposes, a subset of four out of the fifteen \pt intervals is presented. A point for \pp collisions at $\A= 1$, in which the \RAA is unity by definition, is also included in the upper left.}
    \label{fig:fig2}
\end{figure}

Figure~\ref{fig:fig1} shows the measured charged-particle \RAA dependence on $\pt$ for particles with $\abs{\eta}<1$. The \NeNe results are compared to previous CMS measurements~\cite{CMS:2016xef,CMS:2018yyx,CMS:2025bta}. The \NeNe and \OO measurements display a similar \pt dependence, with \NeNe exhibiting a stronger suppression below $\pt \approx 20\GeV$, while converging to comparable \RAA values at higher \pt. The \NeNe measurement provides additional leverage to quantify how rapidly nuclear suppression effects change with the system size relative to \OO. Different models initially predicted varying degrees of separation between these two systems~\cite{CMS-PAS-HIN-25-014}. The observed splitting between \NeNe and \OO hints at a smooth transition with system size. Owing to the current luminosity uncertainties affecting both light ion measurements, particularly the extent to which they are correlated, a definitive statement on their relative suppression magnitude is not yet possible. Data from the larger \XeXe and \PbPb systems show a substantially stronger suppression and a steeper increase of \RAA with \pt. Together with the light ion systems results, this highlights the systematic evolution of the \pt-dependent suppression pattern with collision system size. Overall, the shape of the \RAA as a function of \pt is similar across all systems, with a minimum around 5--7\GeV, a decreasing trend at low \pt, and a rising trend toward higher \pt. In heavy ion systems, the low-\pt behavior is typically attributed to the interplay of \pt-dependent effects such as Cronin enhancement~\cite{Antreasyan:1978cw} and radial flow~\cite{ATLAS:2025ztg, ALICE:2025iud}, while the magnitude and evolution of the suppression at high \pt are generally associated with parton energy loss in the medium.

Figure~\ref{fig:fig2} presents the evolution of the charged-particle \RAA with \cubicA for different AA collision systems and for four representative \pt intervals spanning 9.6--103.6\GeV. Since the nuclear volume grows approximately with \A, \cubicA is proportional to the nuclear radius. Presenting the evolution of the \RAA with \cubicA and in a fixed \pt interval enables a direct model-independent comparison of nuclear suppression effects with the system size. The level of suppression exhibits a monotonic dependence on the system size, with \RAA decreasing as \A (or \cubicA) increases.

\begin{figure*}[ht!]
    \centering
    \includegraphics[width=.9\linewidth]{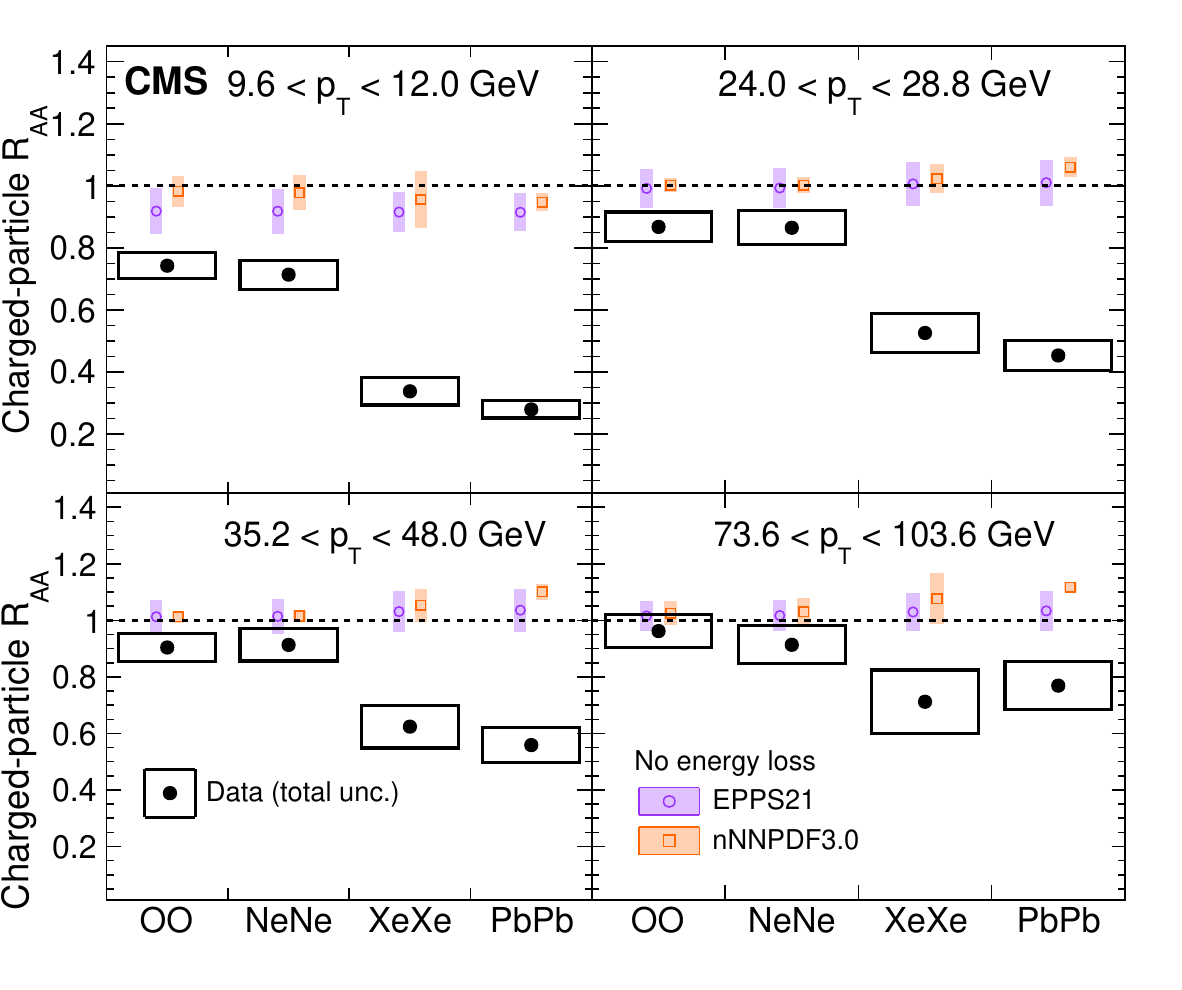}
    \caption{Comparison of the charged-particle \RAA measured in \OO~\cite{CMS:2025bta}, \NeNe, \XeXe~\cite{CMS:2018yyx}, and \PbPb~\cite{CMS:2016xef} collisions with NLO pQCD calculations that incorporate only initial-state effects through nPDFs and do not include parton energy loss~\cite{Huss:2020dwe,Mazeliauskas:2025clt}. The open boxes with black circular markers represent the data. The box height represents the total experimental uncertainty obtained from the quadratic combination of statistical, systematic, and global normalization uncertainties. The purple markers correspond to calculations using the EPPS21 nPDF set ~\cite{Eskola:2021nhw}, while the orange markers represent calculations based on the nNNPDF3.0 nPDF set~\cite{AbdulKhalek:2022fyi}. The vertical uncertainty bands reflect the combined scale-variation and nPDF uncertainties, with the latter providing the dominant contribution.}
    \label{fig:fig3}
\end{figure*}

The measured charged-particle \RAA across systems are compared to a set of theoretical calculations that span baseline nuclear initial-state effects and different implementations of parton energy loss in hot, strongly interacting matter. This cross-system comparison provides more stringent constraints than studies of \RAA versus \pt for individual collision systems, as it reduces the freedom for models to independently describe light and heavy ion collision systems. The calculations discussed here were provided following the first charged-particle \RAA measurement in \OO collisions. A comparison is presented over four representative \pt intervals. A set of theoretical predictions for \NeNe and \OO collisions provided prior to the release of the \OO results is presented in \addmatter{sec:preInitialStages}{Section~6} and, for the OO system, in Ref.~\cite{CMS:2025bta}, respectively. A comparison with the more recent calculations illustrates the difficulty in describing these data with the knowledge available before the \OO experimental results.

\begin{figure*}[ht!]
    \centering
    \includegraphics[width=.9\linewidth]{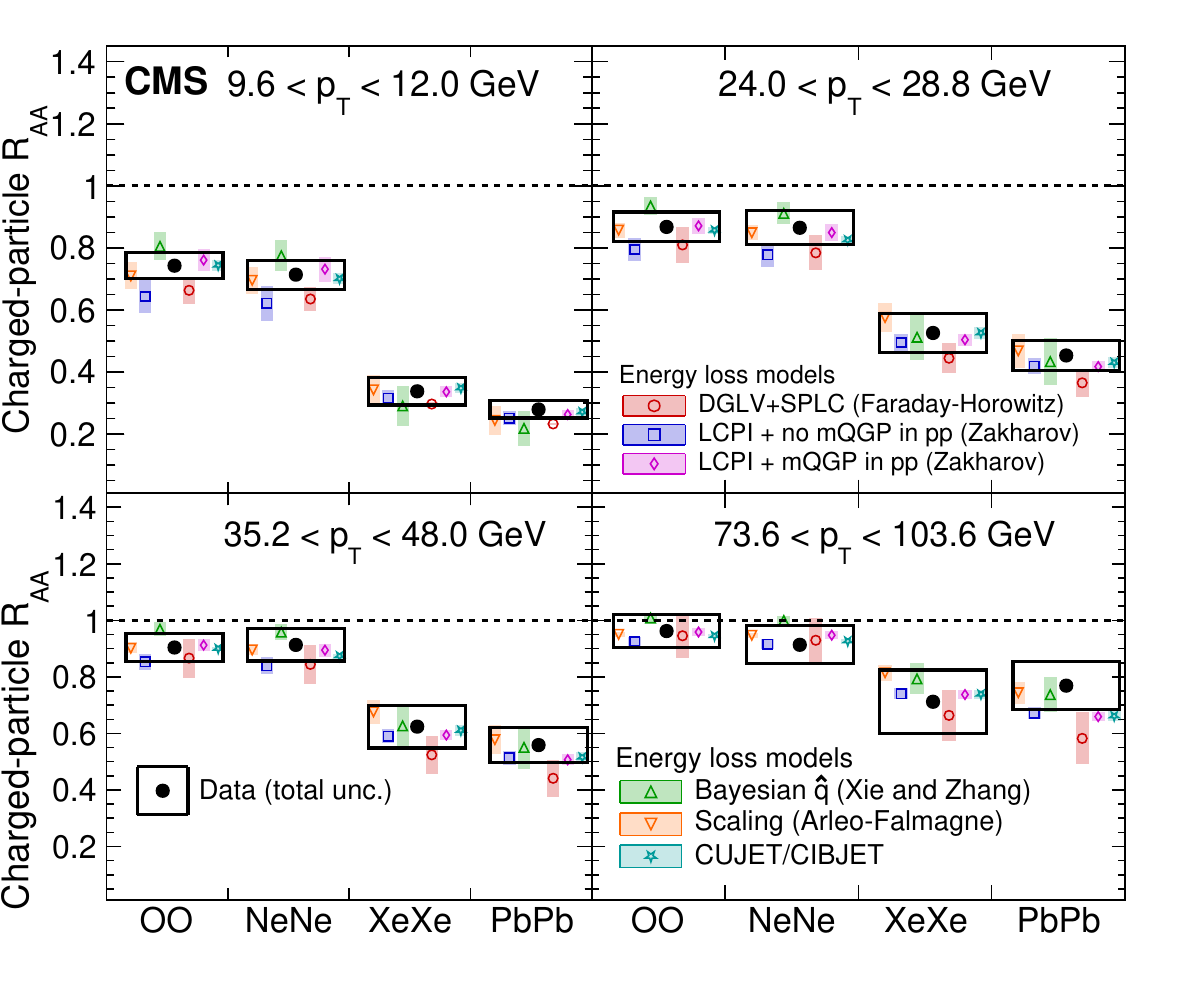}
    \caption{Comparison of the charged-particle \RAA measured in \OO~\cite{CMS:2025bta}, \NeNe, \XeXe~\cite{CMS:2018yyx}, and \PbPb~\cite{CMS:2016xef} collisions with calculations that incorporate parton energy loss effects~\cite{Faraday:2025prr, Faraday:2025pto,Gyulassy:1999zd,Djordjevic:2003zk,Zakharov:2021uza,Arleo:2022shs,Shi:2018izg,Shi:2018lsf,Guo:2025tsf}. The open boxes with black circular markers represent the data, with the box height indicating the total experimental uncertainty obtained from the quadratic combination of statistical, systematic, and global normalization uncertainties. The other colored markers represent theoretical predictions based on different approaches, as described in the text. The uncertainty band is smaller than the marker size for some of the calculations.}
    \label{fig:fig4}
\end{figure*}

Figure~\ref{fig:fig3} compares the measured \RAA to baseline calculations that include only initial-state nuclear effects and no final-state energy loss~\cite{Huss:2020dwe, Mazeliauskas:2025clt}. These calculations are based on next-to-leading order (NLO) perturbative quantum chromodynamics (pQCD) calculations interfaced with modern nuclear parton distribution functions (nPDFs). Calculations using the EPPS21~\cite{Eskola:2021nhw} and nNNPDF3.0~\cite{AbdulKhalek:2022fyi} nPDF sets include the full propagation of nPDF uncertainties, as well as variations of the renormalization and factorization scales. These models predict a modest suppression at low \pt that is largely independent of \A and arises from initial-state nuclear effects alone. As such, these baseline calculations establish a reference against which genuine final-state suppression effects can be isolated. Future analysis of recent pO data sets may further constrain the \A dependence of nPDFs~\cite{Jonas:2026yoz}.

A comparison of theoretical calculations that include parton energy loss versus the data is presented in Fig.~\ref{fig:fig4}. The regime of validity of these begins at approximately $\pt>10\GeV$. The Faraday--Horowitz calculation~\cite{Faraday:2025pto, Faraday:2025prr} employs medium-induced radiative energy loss based on the Djordjevic--Gyulassy--Levai--Vitev (DGLV) formalism~\cite{Gyulassy:1999zd,Djordjevic:2003zk}, supplemented by collisional energy loss contributions and a short path-length correction (SPLC) designed for small collision systems. In this framework, the system-size dependence of \RAA emerges from the interplay between the in-medium path length traversed by hard partons and the evolving temperature profile of the medium. The predicted suppression increases smoothly with increasing \A, and is compatible with the measurements within uncertainties for all collision systems, although some tension is present for the OO and NeNe data points. The calculations from DGLV+SPLC use the same parametrization as in the preliminary results, but they are presented according to the binning employed in this analysis. The measurements are also compared to predictions from the CUJET/CIBJET framework~\cite{Shi:2018izg,Shi:2018lsf}, which is another DGLV-based description of parton energy loss coupled with viscous hydrodynamic modeling of the medium. These calculations incorporate an updated characterization of the medium properties based on a recent Bayesian analysis of the \RAA versus the high-\pt elliptic flow coefficient $v_2$ of heavy ion collisions at RHIC and the LHC~\cite{Guo:2025tsf}. This Bayesian analysis constrained the temperature-dependent chromoelectric and chromomagnetic composition of the QGP, particularly the abundance of chromomagnetic monopoles. The Bayesian analysis does not include light ion collision data. These theoretical calculations are compatible with the data within the uncertainties for these \pt intervals.

Calculations based on the light-cone path integral (LCPI) formalism~\cite{Zakharov:2021uza} include medium-induced radiative energy loss applied to a pQCD baseline with nPDFs. The calculation  shown in Fig.~\ref{fig:fig4} corresponds to model scenarios with or without the formation of a ``mini QGP'' (mQGP) droplet in \pp collisions, leading to different extrapolations toward small \A. The resulting \RAA calculation without QGP formation in \pp exhibits some tension at low \pt with respect to the measured \RAA in light ion collisions, but is compatible for all \pt. The prediction with a QGP formation in \pp is in agreement with the data within the uncertainties.

A complementary approach is provided by calculations that determine the jet transport coefficient \qHat using Bayesian inference constrained by heavy ion data~\cite{Xie:2022ght}. The variable \qHat characterizes the average \pt broadening per unit path length experienced by a high-energy parton traversing the deconfined nuclear medium. In this framework, the space--time evolution of the medium is described using 3+1 dimensional relativistic hydrodynamics~\cite{Pang:2012he}, and energy loss is implemented consistently across collision systems by scaling the temperature and path length distributions experienced by propagating partons. The resulting calculations reproduce the rise of \RAA toward smaller \A, and are compatible with the measurement within the uncertainties.

The results are compared to a model that provides an empirical description of high-\pt hadron suppression in heavy ion collisions~\cite{Arleo:2022shs} based on scaling properties. In this approach, the magnitude and system-size dependence of \RAA are governed by an effective parton energy loss scale inferred from the observed scaling of suppression with collision geometry and particle multiplicity. The scaling parametrization is inspired by the expected relationship between parton energy loss and in-medium path length at leading order in pQCD. Fits of this scaling model to RHIC and LHC data yield an average parton energy loss with an approximately linear dependence of the in-medium path length; light ion collision data are not included in this fit. These scaling-based calculations are in agreement with the data within the uncertainties.

The LCPI and CUJET/CIBJET calculations use charged-particle \pt spectra from \pp collisions calculated at leading order in perturbation theory, whereas DGLV+SPLC and Bayesian use NLO pQCD calculations. Differences in the fragmentation function parametrizations do not substantially change the slope of the predicted charged-particle \pt spectrum in \pp collisions, as summarized in Refs.~\cite{Borsa:2023zxk,dEnterria:2013sgr}. The \RAA observable benefits from substantial cancellations of the renormalization and factorization scale uncertainties and fragmentation function fit uncertainties in the ratio, while remaining sensitive to modifications of particle production arising from nuclear effects.

Overall, models that include parton energy loss naturally generate a monotonic decrease of \RAA with increasing \A, qualitatively consistent with the data. The magnitude and detailed \A dependence vary among models, reflecting differences in the assumptions about medium formation in small and large systems, the path-length dependence of energy loss, and the temperature scaling of the jet transport coefficient \qHat. Although still compatible within the combined experimental and theoretical uncertainties, a subset of the models tend to slightly underestimate the central value of the data, as seen in Fig.~\ref{fig:fig4}. Therefore, the CMS \RAA measurements offer enhanced sensitivity to key aspects of energy loss modeling when considered in aggregate as opposed to individually. This is further supported by the comparison to the theoretical predictions presented prior to the release of the preliminary results, which is described in \addmatter{sec:preInitialStages}{Section~6}. A comparison of the theoretical calculations of \RAA as a function of \pt can be found in \addmatter{sec:RAAvsPt_systemSize}{Section~5}.

\section{Summary}\label{sec:summary}

This Letter presents a systematic study of the system-size dependence of the charged-particle nuclear modification factor (\RAA) versus transverse momentum (\pt). The study is based on a cross-system comparison that minimizes centrality selection biases. For this purpose, the analysis reports the first measurement of the charged-particle \RAA in minimum bias neon-neon (NeNe) collisions at a center-of-mass energy per nucleon pair of 5.36\TeV, based on data collected in 2025 and corresponding to an integrated luminosity of 0.76\nbinv. By comparing existing CMS measurements in oxygen-oxygen, xenon-xenon, and lead-lead collisions with the new \NeNe results, this analysis samples nuclear systems across a broad range of nucleon number values (\A). This enables an evaluation of the role that the system size has on energy loss. The \RAA is measured for charged particles with pseudorapidity $\abs{\eta}<1$ using identical \pt intervals across all collision systems, which allows for a direct comparison of nuclear suppression effects with \A, which characterizes the system size.

The \RAA at a given \pt value is found to monotonically decrease as a function of the cubic root of the nucleon number \cubicA, a quantity that is related to the nuclear radius. The \pt dependences of \RAA in all collision systems have similar qualitative shapes, with a local minimum at about 5--7\GeV. The inclusion of the \NeNe system extends the coverage in system size and enables a more complete mapping of the evolution of suppression effects across collision systems.

Comparisons with theoretical calculations show that baseline perturbative quantum chromodynamics predictions incorporating nuclear parton distribution functions predict only modest modifications relative to $\RAA = 1$, with a weak dependence on \A. In contrast, models that also include parton energy loss reproduce the observed monotonic evolution of \RAA with increasing \A. Differences among these calculations in the predicted \RAA value and its detailed system-size dependence reflect variations in the modeling of medium effects. Overall, models incorporating parton energy loss describe the measurements within experimental and theoretical uncertainties. This study constrains the existence and characteristics of any transition towards a deconfined, hot medium as the system size increases. Hence, this measurement offers guidance for the selection of ion species in future nuclear physics programs at the LHC.

\begin{acknowledgments}

We thank A. Mazeliauskas (nPDF baselines), F. Arleo, G. Falmagne (Scaling), C. Faraday, W. A. Horowitz (DGLV+SPLC), Y. Guo, J. Liao, S. Shi (CUJET/CIBJET), M. Xie, W. Ke, H. Zhang, X.-N. Wang (Bayesian), and B. G. Zakharov (LCPI) for providing theoretical calculations for OO, NeNe, XeXe, and PbPb collision systems using our \pt binning scheme. We also thank I. Kolbe, W. van der Schee, G. Nijs, S. Iqbal, K. Ruhani, and I. Ahmed (Trajectum) for providing NeNe predictions before the results became publicly available.

We congratulate our colleagues in the CERN accelerator departments for the excellent performance of the LHC and thank the technical and administrative staffs at CERN and at other CMS institutes for their contributions to the success of the CMS effort. In addition, we gratefully acknowledge the computing centers and personnel of the Worldwide LHC Computing Grid and other centers for delivering so effectively the computing infrastructure essential to our analyses. Finally, we acknowledge the enduring support for the construction and operation of the LHC, the CMS detector, and the supporting computing infrastructure provided by the following funding agencies: SC (Armenia), BMBWF and FWF (Austria); FNRS and FWO (Belgium); CNPq, CAPES, FAPERJ, FAPERGS, and FAPESP (Brazil); MES and BNSF (Bulgaria); CERN; CAS, MoST, and NSFC (China); MINCIENCIAS (Colombia); MSES and CSF (Croatia); RIF (Cyprus); SENESCYT (Ecuador); ERC PRG and PSG, TARISTU24-TK10 and MoER TK202 (Estonia); Academy of Finland, MEC, and HIP (Finland); CEA and CNRS/IN2P3 (France); SRNSF (Georgia); BMFTR, DFG, and HGF (Germany); GSRI (Greece); NKFIH (Hungary); DAE and DST (India); IPM (Iran); SFI (Ireland); INFN (Italy); MSIT and NRF (Republic of Korea); MES (Latvia); LMTLT (Lithuania); MOE and UM (Malaysia); BUAP, CINVESTAV, CONACYT, LNS, SEP, and UASLP-FAI (Mexico); MOS (Montenegro); MBIE (New Zealand); PAEC (Pakistan); MES, NSC, and NAWA (Poland); FCT (Portugal);  MESTD (Serbia); MICIU/AEI and PCTI (Spain); MOSTR (Sri Lanka); Swiss Funding Agencies (Switzerland); MST (Taipei); MHESI (Thailand); TUBITAK and TENMAK (T\"{u}rkiye); NASU (Ukraine); STFC (United Kingdom); DOE and NSF (USA).

\hyphenation{Rachada-pisek} Individuals have received support from the Marie-Curie program and the European Research Council and Horizon 2020 Grant, contract Nos.\ 675440, 724704, 752730, 758316, 765710, 824093, 101115353, 101002207, 101001205, and COST Action CA16108 (European Union); the Leventis Foundation; the Alfred P.\ Sloan Foundation; the Alexander von Humboldt Foundation; the Science Committee, project no. 22rl-037 (Armenia); the Fonds pour la Formation \`a la Recherche dans l'Industrie et dans l'Agriculture (FRIA) and Fonds voor Wetenschappelijk Onderzoek contract No. 1228724N (Belgium); the Beijing Municipal Science \& Technology Commission, No. Z191100007219010, the Fundamental Research Funds for the Central Universities, the Ministry of Science and Technology of China under Grant No. 2023YFA1605804, the Natural Science Foundation of China under Grant No. 12535004, and USTC Research Funds of the Double First-Class Initiative No.\ YD2030002017 (China); the Ministry of Education, Youth and Sports (MEYS) of the Czech Republic; the Shota Rustaveli National Science Foundation, grant FR-22-985 (Georgia); the Deutsche Forschungsgemeinschaft (DFG), among others, under Germany's Excellence Strategy -- EXC 2121 ``Quantum Universe" -- 390833306, and under project number 400140256 - GRK2497; the Hellenic Foundation for Research and Innovation (HFRI), Project Number 2288 (Greece); the Hungarian Academy of Sciences, the New National Excellence Program - \'UNKP, the NKFIH research grants K 131991, K 133046, K 138136, K 143460, K 143477, K 146913, K 146914, K 147048, 2020-2.2.1-ED-2021-00181, TKP2021-NKTA-64, and 2025-1.1.5-NEMZ\_KI-2025-00004 (Hungary); the Council of Science and Industrial Research, India; ICSC -- National Research Center for High Performance Computing, Big Data and Quantum Computing, FAIR -- Future Artificial Intelligence Research, and CUP I53D23001070006 (Mission 4 Component 1), funded by the NextGenerationEU program (Italy); the Latvian Council of Science; the Ministry of Education and Science, project no. 2022/WK/14, and the National Science Center, contracts Opus 2021/41/B/ST2/01369, 2021/43/B/ST2/01552, 2023/49/B/ST2/03273, and the NAWA contract BPN/PPO/2021/1/00011 (Poland); the Funda\c{c}\~ao para a Ci\^encia e a Tecnologia (Portugal); the National Priorities Research Program by Qatar National Research Fund; MICIU/AEI/10.13039/501100011033, ERDF/EU, ``European Union NextGenerationEU/PRTR", and Programa Severo Ochoa del Principado de Asturias (Spain); the Chulalongkorn Academic into Its 2nd Century Project Advancement Project, the National Science, Research and Innovation Fund program IND\_FF\_68\_369\_2300\_097, and the Program Management Unit for Human Resources \& Institutional Development, Research and Innovation, grant B39G680009 (Thailand); the Eric \& Wendy Schmidt Fund for Strategic Innovation through the CERN Next Generation Triggers project under grant agreement number SIF-2023-004; the Kavli Foundation; the Nvidia Corporation; the SuperMicro Corporation; the Welch Foundation, contract C-1845; and the Weston Havens Foundation (USA).
 
\end{acknowledgments}\section*{Data availability} Release and preservation of data used by the CMS Collaboration as the basis for publications is guided by the  \href{https://doi.org/10.7483/OPENDATA.CMS.1BNU.8V1W}{CMS data preservation, re-use and open access policy}.
\bibliography{auto_generated}
\ifthenelse{\boolean{cms@external}}{}
{\numberwithin{figure}{section}
\appendix

\section{Charged-particle spectrum of the NeNe system}\label{sec:spectrum}

Figure~\ref{fig:spectrum} shows the \pt spectrum of primary charged particles measured in \NeNe collisions at $\sqrtsNN = 5.36\TeV$. The \pp spectrum is the one used in Ref.~\cite{CMS:2025bta}. The spectra are used as an input for the construction of the nuclear modification factor presented in the main text.

\begin{figure}
    \centering
    \includegraphics[width=\cmsFigWidth]{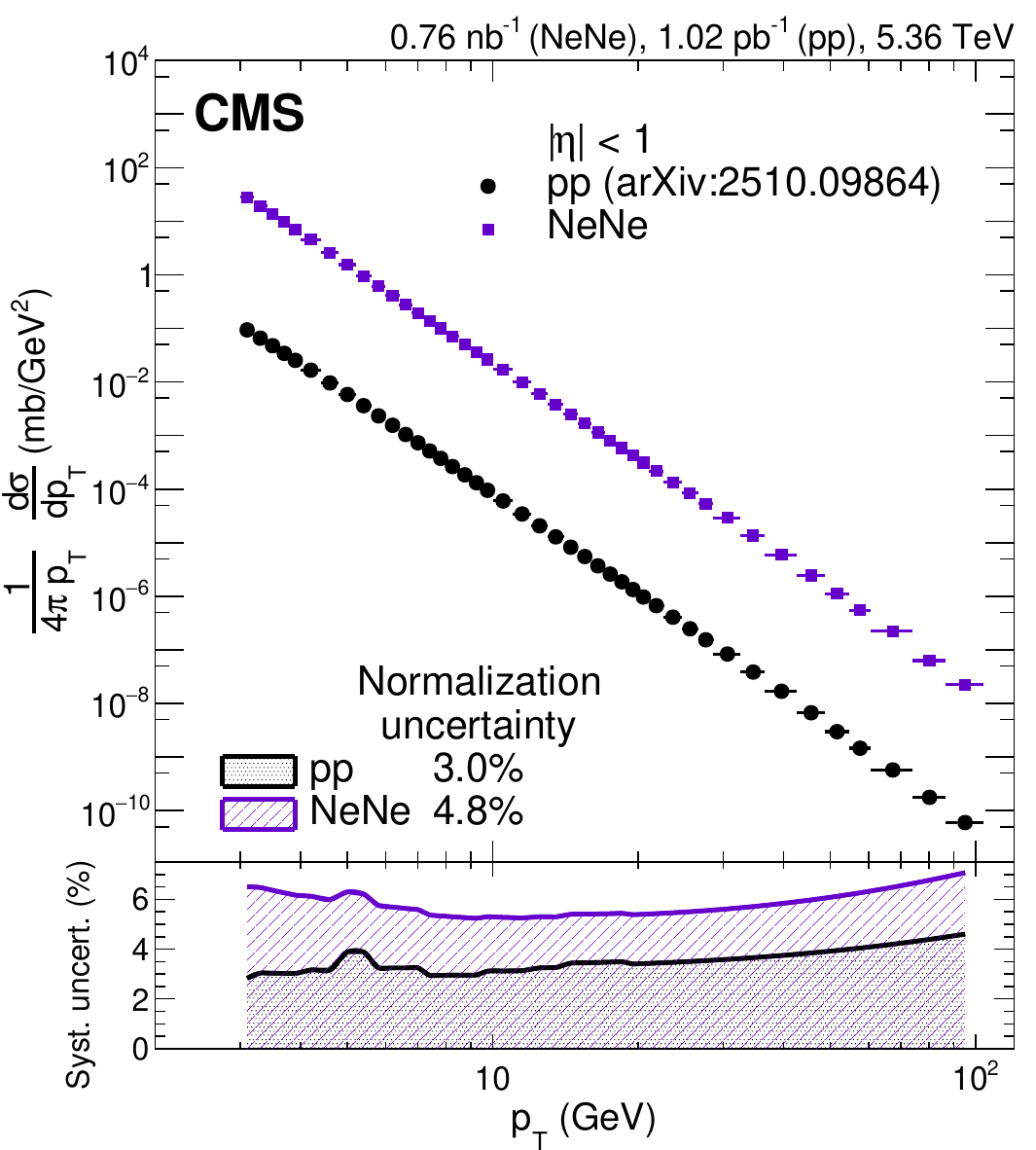}
\caption{Differential invariant cross section spectrum as function of \pt of primary charged particles measured in \NeNe collisions at $\sqrtsNN = 5.36\TeV$ at midrapidity. The \pp spectrum is from Ref.~\cite{CMS:2025bta}. The markers show the average cross section across the entire bin width, not the cross section value at the center of each bin. Statistical uncertainties are smaller than the markers.
}
    \label{fig:spectrum}
\end{figure}

\section{Bin merging procedure for PbPb \texorpdfstring{\RAA}{RAA} }\label{app:binning}

The \pt-binning of the \PbPb \RAA measurement is finer than the \XeXe binning at high \pt. The \PbPb \RAA bins are merged according to the binning scheme of the \XeXe \RAA measurement~\cite{CMS:2018yyx}. The bin edges from the \XeXe measurement are shared with the \PbPb measurement bins. To match the \pt binning constraints of the \XeXe analysis, the two bins of the \PbPb \RAA having ranges 35.2--41.6 and 41.6--48.0\GeV were merged into a single 35.2--48.0\GeV bin, and the two bins having ranges 48.0--60.8 and 60.8--73.6\GeV were merged into 48.0--73.6\GeV. Bin merging was performed at the level of the \PbPb and \pp~spectra to account for the in-bin \pt shape of the individual \PbPb and \pp distributions used to compute \RAA.

Bins within the \PbPb and \pp spectra were merged separately using the HEPData entry associated with Ref.~\cite{CMS:2016xef}. In this procedure, central values were combined as a \pt-weighted average. This is because the spectra were reported as invariant cross sections
with a prefactor of $1/\pt$, which must be removed before the combination of yields in the bin merging procedure. Statistical uncertainties of the bins were added in quadrature while the systematic uncertainties were combined linearly and averaged for each merged bin, assuming fully correlated bin-to-bin uncertainties. Both uncertainties were weighted by the \pt and bin widths of the constituent bins. The formulas below summarize the bin-merging procedure for the cross section ($\sigma$) corresponding to a merged bin in the numerator of the charged-particle \RAA in \PbPb collisions.

Consider the set of original \pt bins indexed by \bini, which are combined  into a single merged bin. For each original bin \bini, \sigBin denotes the  measured cross section, \ptwidthi denotes the bin width, and \ptcenteri denotes  the bin center. The quantities \sigBinStat and \sigBinSyst denote the
statistical and systematic uncertainties, respectively. The merged bin is  characterized by a total width \ptwidthMerged and a representative center  \ptcenterMerged. The central value of the merged cross section is obtained as
\begin{equation}
\sigMerged =
\frac{
\sum_{\bini}
\sigBin \ptwidthi \ptcenteri
}{
\ptwidthMerged \ptcenterMerged
}.
\end{equation}
The statistical uncertainty of the merged bin, assuming the statistical
uncertainties of the original bins are uncorrelated, is computed as
\begin{equation}
\sigMergedStat =
\frac{
\left[
\sum_{\bini}
\left(
\sigBinStat \ptwidthi \ptcenteri
\right)^2
\right]^{1/2}
}{
\ptwidthMerged \ptcenterMerged
}.
\end{equation}
The systematic uncertainty, treated as fully correlated across the merged bins, is evaluated as
\begin{equation}
\sigMergedSyst =
\frac{
\sum_{\bini}
\sigBinSyst  \ptwidthi  \ptcenteri
}{
\ptwidthMerged  \ptcenterMerged
}.
\end{equation}
The normalization uncertainty remains unaffected by the bin-merging procedure. The same bin merging procedure was employed for the denominator of the \pp reference system. Although the main focus of this analysis is on symmetric AA collisions, for completeness, we also repeated the procedure for the \pPb system using the inputs presented in the same HEPData record.

\section{Charged-particle \texorpdfstring{\RAA}{RAA} with nominal \texorpdfstring{\pt}{pT} binning}\label{app:nominal}

Figure~\ref{fig:nominalRAA} shows the charged-particle nuclear modification factor \RAA measured with the nominal \pt binning used for each collision system. For \OO, \NeNe, and \PbPb, the binning is finer than the coarser binning used for the common system comparison presented in Fig.~\ifthenelse{\boolean{cms@external}}{Fig. 4 in the main documents}{\ref{fig:fig4}}.

\begin{figure}
    \centering
    \includegraphics[width=\cmsFigWidth]{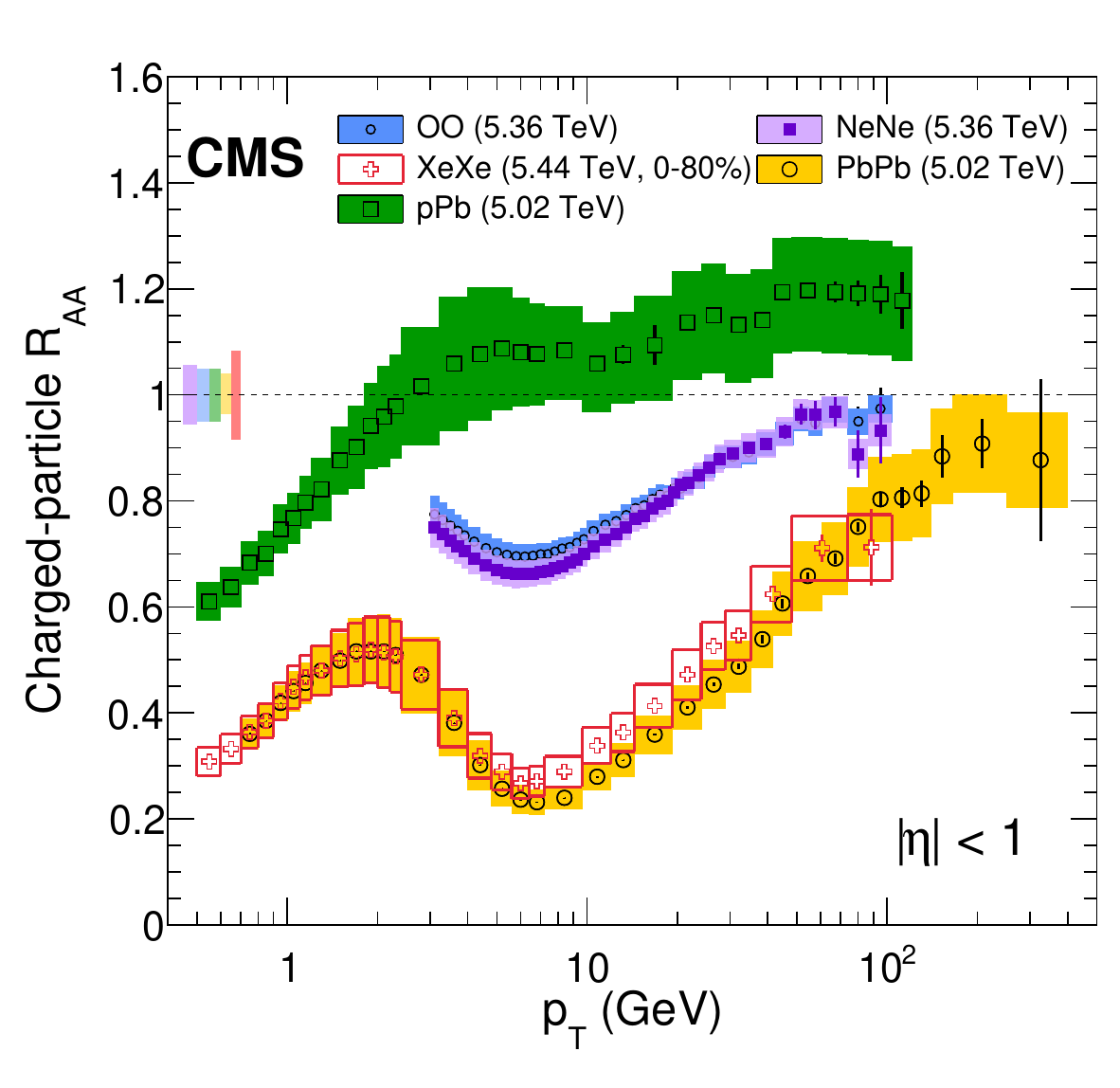}
\caption{Charged-particle nuclear modification factor \RAA measured using the nominal \pt binning for \OO, \NeNe, \XeXe, \PbPb, and \pPb collisions. The vertical bands around the markers indicate the total experimental uncertainties, while the vertical error bars show the statistical uncertainties. The boxes on the left, centered at unity, represent the relative normalization uncertainty for each collision system.}
    \label{fig:nominalRAA}
\end{figure}

\section{Charged-particle \texorpdfstring{\RAA}{RAA} with coarser \texorpdfstring{\pt}{pT} binning, including the pPb system}

Figure~\ref{fig:coarseRAA} shows the charged-particle nuclear modification factor \RAA measured with a coarser \pt binning. The \pPb system is included for comparison. The results are fully corrected and correspond to those used in the main analysis.

\begin{figure}
    \centering    \includegraphics[width=\cmsFigWidth]{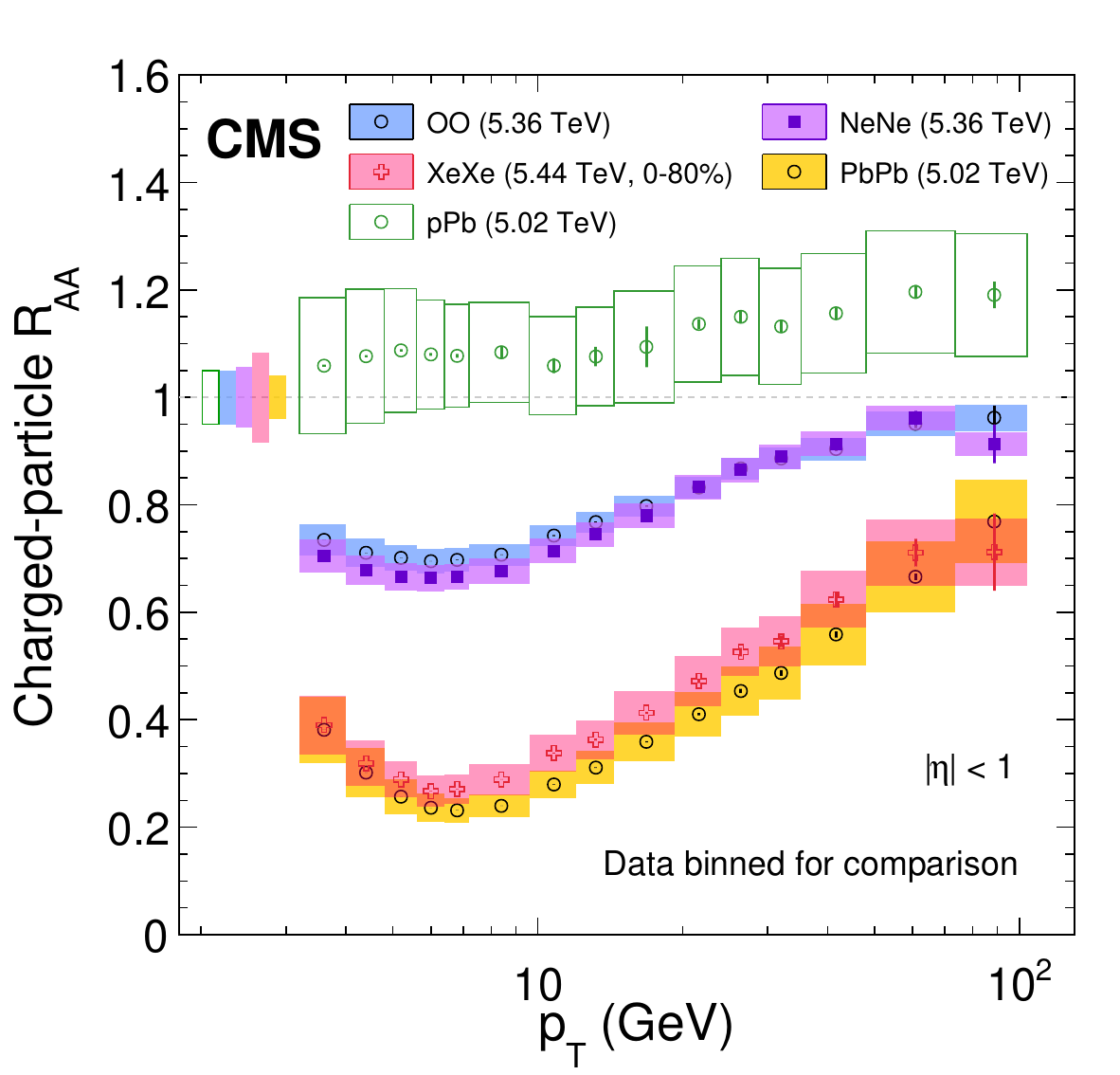}
\caption{Charged-particle nuclear modification factor \RAA measured using the common \pt binning scheme for \OO, \NeNe, \XeXe, \PbPb, and \pPb collisions. The vertical bands around the markers indicate the total experimental uncertainties, while the vertical error bars show the statistical uncertainties. The boxes on the left, centered at unity, represent the relative normalization uncertainty for each collision system.}
    \label{fig:coarseRAA}
\end{figure}

\section{Theoretical calculations for \texorpdfstring{\RAA}{RAA} as functions of \texorpdfstring{\pt}{pT} for all systems}\label{sec:RAAvsPt_systemSize}

Figures~\ref{fig:RAA_vs_pT_noQuenching} and \ref{fig:RAA_vs_pT_quenching} present theoretical calculations of the charged-particle \RAA versus \pt for minimum bias \OO and \NeNe collisions, \XeXe collisions in the 0--80\% centrality interval, and minimum bias \PbPb collisions, compared to the data in the same intervals. These are the same theoretical calculations presented in the main body of the Letter. Figure~\ref{fig:RAA_vs_pT_noQuenching} shows calculations that include only initial-state nuclear effects and no final-state energy loss, whereas Fig.~\ref{fig:RAA_vs_pT_quenching} presents calculations that incorporate parton energy loss in the hot and dense medium produced in AA collisions. The comparison highlights the substantial additional suppression observed in the data when final-state interactions are present, particularly at high \pt, and illustrates the limited system-size dependence expected from initial-state effects alone over the range of collision systems and \pt considered.

\begin{figure*}
    \centering
    \includegraphics[width=.9\linewidth]{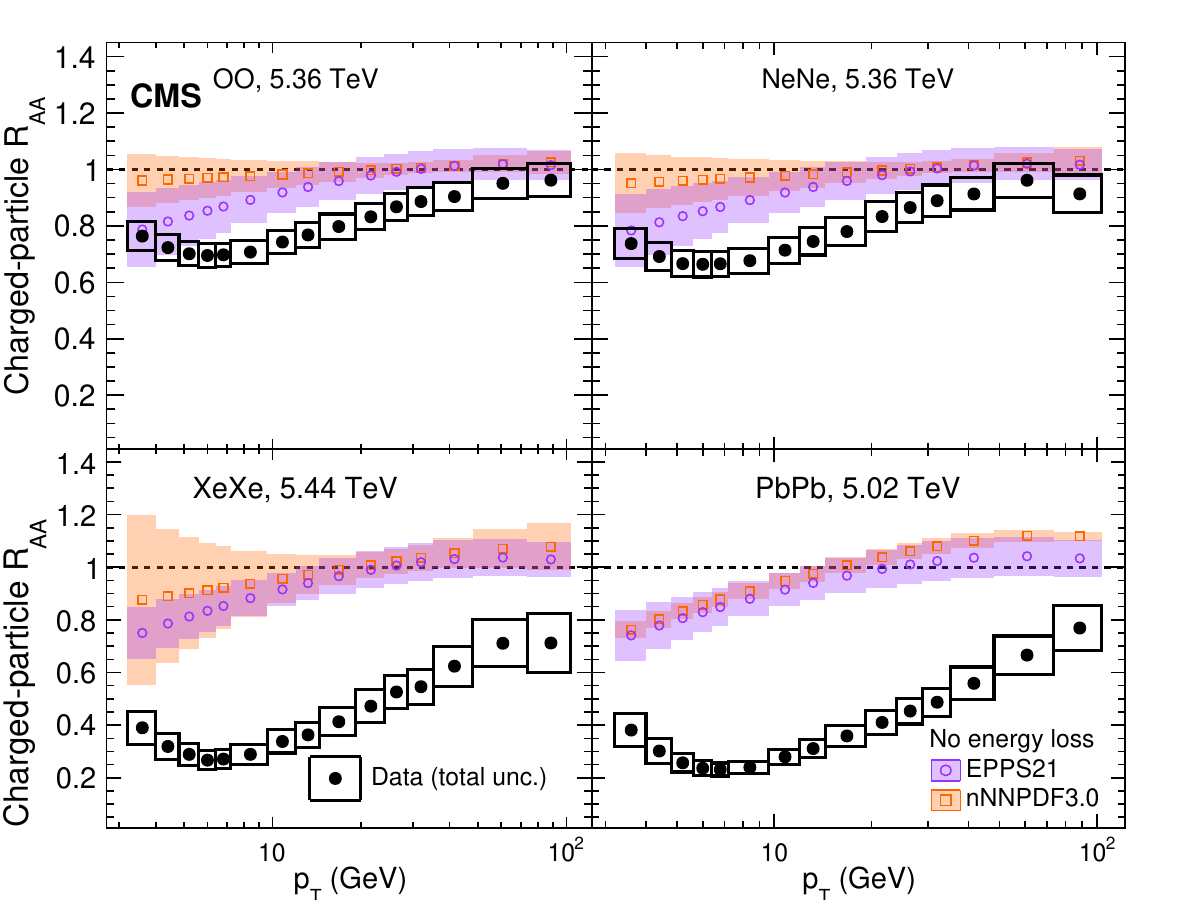}
    \caption{Charged-particle \RAA as a function of \pt in minimum bias \OO ($\sqrtsNN=5.36\TeV$), minimum bias \NeNe (5.36\TeV), \XeXe (0--80\%, 5.44\TeV), and minimum bias \PbPb (5.02\TeV) collisions. The measurements are compared with theoretical calculations without final-state energy loss~\cite{Huss:2020dwe,Mazeliauskas:2025clt}, illustrating the expected size of initial-state nuclear effects. The purple markers correspond to calculations using the EPPS21 nPDF set ~\cite{Eskola:2021nhw}, while the orange markers represent calculations based on the nNNPDF3.0 nPDF set~\cite{AbdulKhalek:2022fyi}. Details of the calculations are discussed in the main body of the text.}
    \label{fig:RAA_vs_pT_noQuenching}
\end{figure*}

\begin{figure*}
    \centering
    \includegraphics[width=.9\linewidth]{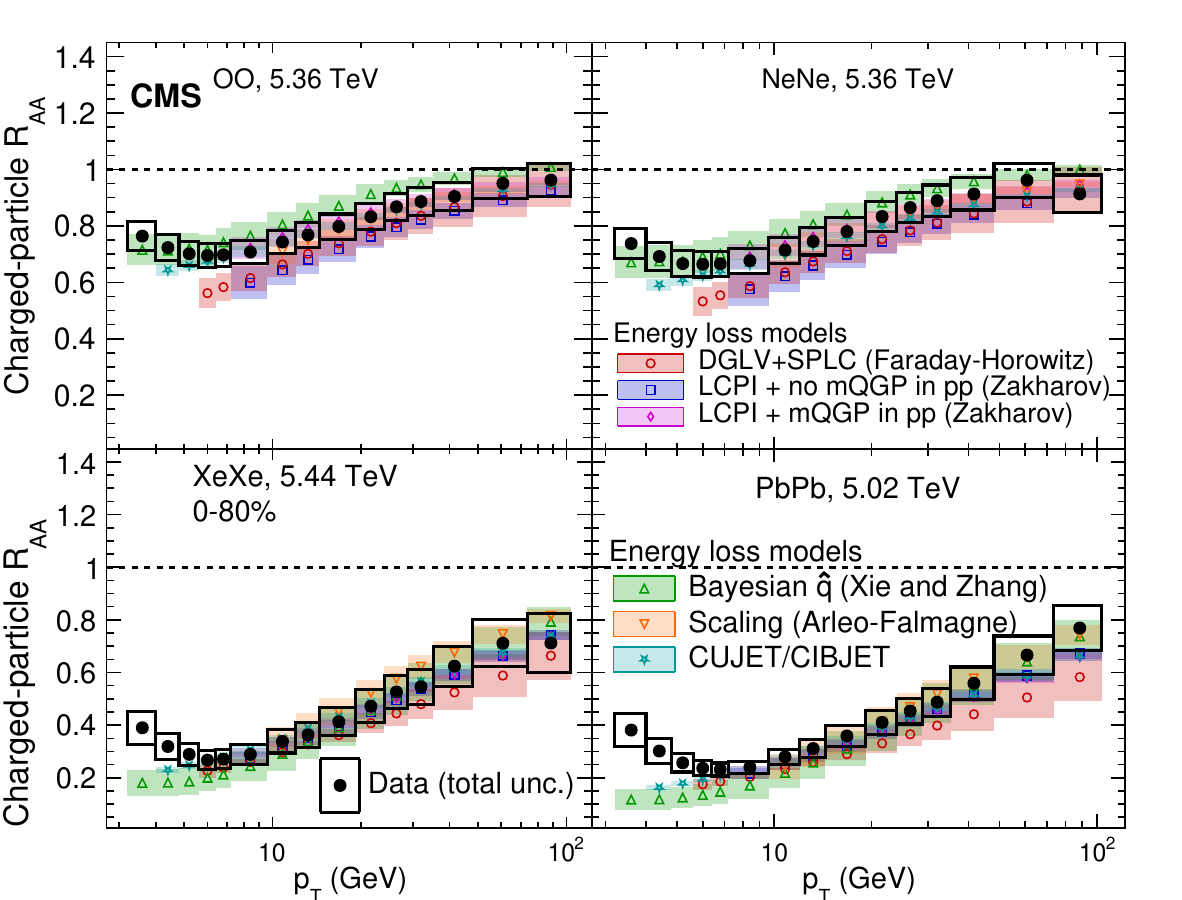}
    \caption{Charged-particle \RAA as a function of \pt in minimum bias \OO ($\sqrtsNN=5.36\TeV$), minimum bias \NeNe (5.36\TeV), \XeXe (0--80\%, 5.44\TeV), and minimum bias \PbPb (5.02\TeV) collisions. The measurements are compared with theoretical calculations that include parton energy loss in the medium~\cite{Faraday:2025prr, Faraday:2025pto,Gyulassy:1999zd,Djordjevic:2003zk,Zakharov:2021uza,Arleo:2022shs,Shi:2018izg,Shi:2018lsf,Guo:2025tsf}. Details of the calculations are discussed in the main body of the text.}
    \label{fig:RAA_vs_pT_quenching}
\end{figure*}

\section{Theoretical predictions for NeNe \texorpdfstring{\RAA}{RAA} versus \texorpdfstring{\pt}{pT} provided prior to preliminary results}\label{sec:preInitialStages}

Figure~\ref{fig:NeNeTheory} presents theoretical predictions provided prior to the release of the preliminary results of the charged-particle \RAA measurements in \NeNe and \OO collisions.

\begin{figure}
    \centering
    \includegraphics[width=\cmsFigWidth]{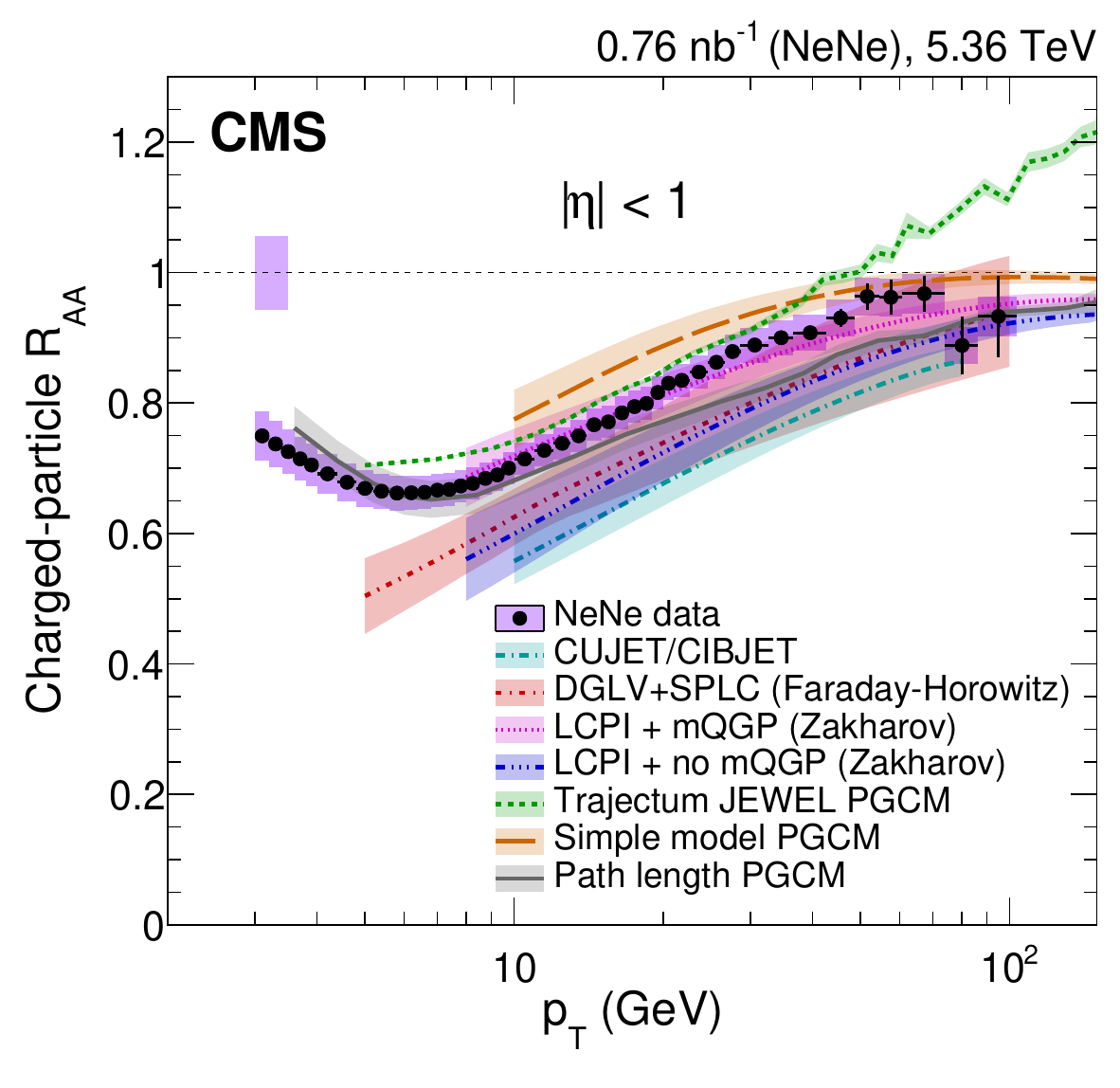}
    \caption{Theoretical predictions for the charged-particle \RAA in \NeNe collisions compared to the measured \NeNe \RAA~\cite{Eskola:2009uj,Zakharov:2021uza,Faraday:2025pto,Shi:2018izg,Shi:2018lsf,Djordjevic:2003zk,Gyulassy:1999zd,vanderSchee:2025hoe,Frosini:2021tuj,Yao:2018qjv,Yao:2019rck,Frosini:2021fjf,Frosini:2021sxj,Frosini:2021ddm,Huss:2020whe,Zapp:2008gi,Zapp:2012ak,Zapp:2013vla,Kolbe:2023rsq,Nijs:2020ors,Nijs:2020roc,Eskola:2021nhw,vanderSchee:2023uii}, provided prior to the release of the preliminary \NeNe and \OO results.}
    \label{fig:NeNeTheory}
\end{figure}

Leading-order pQCD calculations based on the LCPI formalism that include nuclear modifications of the parton distribution functions via the EPS09 set~\cite{Eskola:2009uj}, both assuming and neglecting the formation of a QGP droplet in the \pp reference system (with or without a ``mini QGP'', mQGP)~\cite{Zakharov:2021uza}, reproduce the measured \pt dependence and overall suppression magnitude within experimental uncertainties. A pQCD-based calculation implementing the DGLV opacity expansion with small-system corrections (DGLV+SLPC)~\cite{Faraday:2025pto} provides a similarly good description of the data across the measured \pt range.

Predictions from the CUJET/CIBJET framework~\cite{Shi:2018izg,Shi:2018lsf}, which couples viscous hydrodynamics to a DGLV energy loss formalism~\cite{Djordjevic:2003zk,Gyulassy:1999zd} but does not include nPDF effects, yield systematically lower \RAA values than observed in the data. The theoretical uncertainties in this calculation arise primarily from variations in the strong coupling constant.

The data are further compared to models described in Ref.~\cite{vanderSchee:2025hoe}, which incorporate nuclear structure inputs from the projected generator coordinate method (PGCM)~\cite{Frosini:2021tuj,Yao:2018qjv,Yao:2019rck,Frosini:2021fjf,Frosini:2021sxj,Frosini:2021ddm}. Theoretical predictions using an alternative set of nuclear structure inputs from  nuclear lattice effective field theory ~\cite{Lee:2008fa,Lahde:2019npb,lee_recent_2020} were also provided and yield very similar \RAA values as PGCM-based predictions~\cite{vanderSchee:2025hoe}. An implementation based on the Baier-Dokshitzer-Mueller-Peign\'{e}-Schiff--Zakharov formalism~\cite{Baier:1996kr,Baier:1998yf,Baier:2001yt,Zakharov:1997uu}, referred to as the ``Simple model''~\cite{vanderSchee:2025hoe,Huss:2020whe}, tends to overestimate the measured \RAA, with the discrepancy decreasing toward higher \pt. Predictions obtained by coupling the JEWEL event generator~\cite{Zapp:2008gi,Zapp:2012ak,Zapp:2013vla,Kolbe:2023rsq} to the \textsc{Trajectum} hydrodynamic framework~\cite{Nijs:2020ors,Nijs:2020roc}, including nPDF effects via the EPPS21 set~\cite{Eskola:2021nhw}, overestimate the measured \RAA and predict values exceeding unity at \pt of approximately 50\GeV. An alternative energy loss model that parametrizes the suppression through a path-length- and \pt-dependent energy loss~\cite{vanderSchee:2023uii}, combined with the hydrodynamic description of \textsc{Trajectum}, qualitatively reproduces key features of the \NeNe measurement, including the presence of a local minimum in \RAA, although the minimum occurs at a slightly higher \pt than observed in the data.
}\cleardoublepage \section{The CMS Collaboration \label{app:collab}}\begin{sloppypar}\hyphenpenalty=5000\widowpenalty=500\clubpenalty=5000\cmsinstitute{Yerevan Physics Institute, Yerevan, Armenia}
{\tolerance=6000
A.~Belyaev\cmsorcid{0000-0003-1692-1173}, A.~Gevorgyan\cmsorcid{0000-0003-2751-9489}, A.~Hayrapetyan, A.~Tumasyan\cmsAuthorMark{1}\cmsorcid{0009-0000-0684-6742}
\par}
\cmsinstitute{Marietta Blau Institute for Particle Physics, Vienna, Austria}
{\tolerance=6000
W.~Adam\cmsorcid{0000-0001-9099-4341}, L.~Benato\cmsorcid{0000-0001-5135-7489}, T.~Bergauer\cmsorcid{0000-0002-5786-0293}, M.~Dragicevic\cmsorcid{0000-0003-1967-6783}, S.~Gundacker\cmsorcid{0000-0003-2087-3266}, A.K.~Guven\cmsorcid{0009-0004-5670-5138}, P.S.~Hussain\cmsorcid{0000-0002-4825-5278}, M.~Jeitler\cmsAuthorMark{2}\cmsorcid{0000-0002-5141-9560}, N.~Krammer\cmsorcid{0000-0002-0548-0985}, A.~Li\cmsorcid{0000-0002-4547-116X}, D.~Liko\cmsorcid{0000-0002-3380-473X}, M.~Matthewman, J.~Schieck\cmsAuthorMark{2}\cmsorcid{0000-0002-1058-8093}, R.~Sch\"{o}fbeck\cmsAuthorMark{2}\cmsorcid{0000-0002-2332-8784}, M.~Shooshtari\cmsorcid{0009-0004-8882-4887}, M.~Sonawane\cmsorcid{0000-0003-0510-7010}, N.~Van~Den~Bossche\cmsorcid{0000-0003-2973-4991}, W.~Waltenberger\cmsorcid{0000-0002-6215-7228}, C.-E.~Wulz\cmsAuthorMark{2}\cmsorcid{0000-0001-9226-5812}
% ...
\par}
\vskip\cmsinstskip
\dag:~Deceased\\
$^{1}$Also at Yerevan State University, Yerevan, Armenia\\
$^{2}$Also at TU Wien, Vienna, Austria\\
% ...
\end{sloppypar}
\end{document}